# Cosmic Fine Tuning and the Multiverse Hypothesis

Colin S. Coleman
Defence Science and Technology Organisation
Edinburgh, Australia 5111

## ABSTRACT

The observable universe is necessarily hospitable for life. There are indications, however, that the laws of physics and cosmological parameters need not take the form and values observed, and if they were slightly different life could not exist. A common approach to this fine tuning problem is to propose a cosmos with an ensemble of domains, mostly inhospitable for life. A Bayesian method is used to show that this hypothesis is more credible than a homogeneous fine tuned universe. This conclusion is straightforward for a finite ensemble, but can be extended to an infinite ensemble by applying a formulation of the Principle of Mediocrity.

## 1. Introduction

Cosmology involves several challenges that distinguish it from other disciplines. It concerns domains that are inaccessible to observation, it confronts the notion of infinity as an expression of reality, and it is strongly influenced by observer selection effects. Observer selection means the apparent properties of the universe are selected in that they must be such as to permit the emergence of observers. What is observed, therefore, may be very unrepresentative of what exists.

Despite strong agreement between theory and observation in modern cosmology, critical gaps in knowledge remain. These include the initiation of the big bang, the nature of dark matter and dark energy, and whether the universe is finite or infinite. Also, if the properties of the universe were slightly different, no life could exist.

This last issue is known as the fine tuning problem, which was first noted in the coincidences of stellar nucleosynthesis (Hoyle, 1965) and subsequently in a broader cosmological context Carr & Rees, 1979). It concerns the fact that the physical laws and cosmological parameters are not constrained to fall within the range that permits observers to emerge. If these laws apply everywhere, this circumstance requires explanation.

A common approach to the fine tuning problem is to propose a cosmos that is far larger than that observed, much of it being unsuitable for observers. In this so called multiverse, the local region has properties that permit observers and so appears fine tuned. The entire multiverse, however, is deemed not to be fine tuned.

The multiverse concept continues an enduring trend in cosmology. Previous anthropocentric views (geocentric, heliocentric, galactocenteric) all proved to be false. Multiverse cosmology takes this a step further by removing any central status from the entire Hubble volume. The properties of domains beyond the horizon may differ, but all observers must be located in regions where the properties permit observers to exist.

Multiverse cosmology has been criticized for being unfalsifiable, and hence unscientific. It is tempting to dismiss the concept on these grounds, but a lack of falsifiability does not mean an idea is false. To dismiss unfalsifiable hypotheses is to impose another anthropocentric condition on the universe. Falsifiable theories are naturally desirable because their credence increases in response to ever more stringent attempts to reject them. If this is not possible, however, an alternative approach is required.

Multiverse cosmologies are addressed here by using Bayes' theorem to determine their credence relative to a homogeneous fine tuned universe. Key to this approach is the treatment of observer selection effects by the Self Sampling Assumption (SSA) (Bostrom, 2002). This requires that the observer be regarded as a random member of the set of all observers. Multiverse cosmology is shown to be more credible than a homogeneous fine tuned universe, and the strength of this inference is given by the degree of fine tuning.

This result is direct for a finite multiverse, but less so for an infinite ensemble. One problem is that observers may occur due to random fluctuations of matter, albeit with extremely low probability. Such 'freak' observers are not confined to domains that permit observers to emerge naturally, and may be more numerous than ntural observers if observer-supporting universes are sufficiently rare. This makes it impossible to treat observer selection effects by applying SSA.

To address this problem a Principle of Mediocrity is proposed whereby a section of the the multiverse containing the observer is unexceptional if viewed on a sufficiently large scale. A weak form of this principle is used to show that observer-supporting members of the ensemble comprise an infinite subset with non-zero measure. This allows freak observers to be neglected, and ensures the ratio of prior probabilities is non-singular, thereby extending the result to the case of an infinite multiverse.

## 2. Multiple Universe Cosmologies

The multiverse concept is not founded solely upon the fine tuning problem. Inflation cosmology and quantum theory provide a strong basis for a heterogenous ensemble of universes (Linde, 2007). String theory, for example, suggests the existence of a vast number of domains with different physical laws depending on the geometry of the compact dimensions.

A four level taxonomy of multiverse cosmologies has been proposed (Tegmark, 2007). The first level includes the infinite ergodic universe, which is a consequence of chaotic inflation. This comprises many Hubble volumes with all possible initial conditions but homogeneous laws.

The level 2 multiverse envisages many replications of level 1, viewed as 'bubbles' in an inflating medium where inflation has ceased. Spontaneous symmetry breaking results in different laws and cosmological parameters in each such bubble.

Level 3 is motivated by the many-worlds interpretation of quantum mechanics. Every quantum state of the universe exists simultaneously, rather than colapsing to a single state upon making an observation. This adds no content beyond level 2, the main distinction being that the ensemble exists in an infinite dimensional Hilbert space rather than real physical space.

Finally, the level 4 multiverse is motivated by a principle of mathematical democracy, in which all possible mathematical structures exist in reality. This derives from philosophical considerations, and serves to close the multiverse hierarchy.

The key feature of all multiverse models is the variation of physical laws. If the laws were homogeneous, with variation only in initial conditions, the concept offers no resolution to the fine tuning problem. This is the case for the infinite ergodic universe at level 1. In general, however, the laws may vary over distances larger than the Hubble scale to provide the necessary inhomogeneity. The level 2 multiverse envisages disjoint domains with different physical laws. For the purpose of resolving the fine tuning problem, it is of no concern whether the variation is continuous or discrete.

## 3. Bayesian Inference and the Self Sampling Assumption

Science progresses by enhancing the credence of a hypothesis in response to evidence. When an observer's existence depends on the hypothesis in question, observer selection effects must be taken into account. In cosmology this is expressed by the Anthropic Principle (AP) (Carter, 1974; Barrow & Tipler 1986) originally stated by Carter as *...what we can expect to observe must be restricted by the conditions necessary for our presence as observers.*

Various forms of AP have been proposed. Some have been criticized for being tautological or offering no prescription for application. These concerns have been addressed by re-casting AP as the Self Sampling Assumption (SSA) stated as follows: *One should reason as if one were a random sample from the set of all observers in one's reference class* (Bostrom, 2002).

This approach avoids tautology and indicates a methodology for expressing the probabilistic connection between theory and observation. It provides a means to determine observational consequences given theories about the distribution of observers. The intent of SSA is to treat an observation as a random element of the set, or reference class, of all comparable observations. This has been formalized by Bostram in an observation equation giving the probability of a hypothesis *h* as a consequence of evidence *e*.

No attempt is made here to apply the observation equation directly to specific multiverse models. Current prescriptions lack the detail required to calculate the distribution of observers and construct the observer reference class. Instead, Bayes' theorem and SSA are used to determine the credence of a composite multiverse hypothesis relative to a fine tuned universe with homogeneous properties that permit observers.

## 4. A Thought Experiment

To illustrate the rationale, consider an observer with no knowledge of the outside world. She finds herself in a room with one feature, a four digit number $N$ equal the year of her birth. What can be deduced about the nature of such rooms?

Two alternate hypotheses may be considered. In the first hypothesis ($H_1$) there are many rooms, each with the same four digit number. An alternate hypothesis ($H_2$) has many rooms with an ensemble of different numbers. The evidence $E$ is that this room has a number equal to the observer's birth year.

Assuming no causal relationship between room numbers and observers, it would be a coincidence for all rooms to have a feature specific to the observer. The ratio of prior probabilities $P(H_1) / P(H_2) = 10^{-4}$ if in $H_2$ all four digit numbers are equally probable. The conditional probability of $E$ under each hypothesis is $P(E/H_1) = 1$ and $P(E/H_2) = 10^{-4}$. Then by Bayes' theorem:

$$\frac{P(H_1/E)}{P(H_2/E)} = \frac{P(E/H_1)P(H_1)}{P(E/H_2)P(H_2)} = \frac{10^{-4}}{10^{-4}} = 1$$

Thus the evidence $E$ offers no reason to favor either hypothesis.

With no causal relationship between rooms and observers, no observer selection effect is involved. Such an effect may be introduced by specifying that an occupant can survive only in a room with the property that its number is equal to the year of her birth. All observers must then find themselves in such a room.

In this case it remains a coincidence that all rooms have a feature specific to the observer, so $P(H_1) / P(H_2) = 10^{-4}$ as before. Applying SSA with a reference class of all observers in all rooms, the conditional probability of $E$ under $H_1$ or $H_2$ is unity, and hence:

$$\frac{P(H_1/E)}{P(H_2/E)} = \frac{P(E/H_1)P(H_1)}{P(E/H_2)P(H_2)} = 10^{-4}$$

When observers are constrained to occupy rooms that are amenable to their existence, therefore, the ensemble hypothesis is favored over one in which all rooms are identical.

## 5. The Multiverse Hypotheses

The observable universe appears to be fine tuned. There is dispute over the degree to which this is the case, but it is widely accepted that at least some cosmological parameters are not constrained to take their observed values, and the allowed range of values is much greater than that which permits the emergence of observers. Fine tuning may be quantified by a parameter $F$, the probability of selecting by chance properties that permit the emergence of observers.

Now consider cosmological hypotheses $H_1$, $H_2$ and evidence $E$ as follows:

$H_1$: Physical laws and cosmological parameters are homogeneous and permit observers.

$H_2$: Physical laws and cosmological parameters occur in a *finite* ensemble $M$ with a non-empty subset $S$ that permit observers.

$E$: The observed universe permits observers

Following the method of the thought experiment, cosmology $H_1$ is fine tuned but $H_2$ is not, and the ratio of prior probabilities $P(H_1) / P(H_2) = F$. Applying SSA with the reference class of all observers in the ensemble, the conditional probability of $E$ under $H_1$ or $H_2$ is unity, since in $H_2$ all observers are in an element of $S$. Then by Bayes' theorem:

$$\frac{P(H_1/E)}{P(H_2/E)} = \frac{P(E/H_1)P(H_1)}{P(E/H_2)P(H_2)} = F$$

If the observable universe is fine tuned with $F \ll 1$, the evidence that it permits observers strongly favors the ensemble hypothesis $H_2$, and the greater the fine tuning the greater the inference favoring $H_2$ over $H_1$. This argument holds for a finite ensemble because the existence of the observable universe means that $S$ is not empty, and the ratio of prior probabilities is not singular.

## 6. The Infinite Multiverse

There is no reason to expect the multiverse ensemble to be finite. Indeed a countably infinite ensemble is a more natural model. The previous analysis does not hold for an infinite ensemble for two reasons. Firstly, if the set of observer-supporting elements is finite, or infinite with zero measure, the ratio of prior probabilities is singular and the relative credence of the two hypotheses is indeterminate. Secondly, freak observers may occur in any element of $M$ that contains matter, which may be more numerous than those that permit observers to emerge by normal processes. Consequently, the assumption that all observers occur in observer-permitting elements of the ensemble is false.

To deal with the infinite ensemble it is necessary to show that the measure of $S$ is non-zero. With no established theory of universe formation, however, it is not possible to determine the probability that a given element of $M$ has properties that permit observers, and hence no way to determine the measure of $S$.

The Principle of Mediocrity is commonly accepted in cosmology. This holds broadly that a region containing the observer is not special if viewed on a sufficiently large scale. Stated differently, the local region is statistically unexceptional within the set of all similar regions at large scales. This principle may also be applied to the multiverse. The observable universe is special in that it has properties that permit observers, but a large finite section of the multiverse, containing the observable universe, should be unexceptional within the set of all similar sections.

Consider a partition of the multiverse into a series of disjoint finite domains $P_i$. Each $P_i$ contains $m_i$ elements of $M$ and $s_i \leq m_i$ elements of $S$. The proportion of observer-supporting elements $\rho_i = s_i/m_i$ satisfies a distribution function $F(\rho_i) = P(\rho_j \leq \rho_i)$; $j \neq i$. The local universe may be assumed to be in $P_1$ and the Principle of Mediocrity defined as follows:

**For partition $P_i$ and variable $\rho_i$ with distribution $F(\rho_i)$, $F(\rho_1) \neq 0$ or $1$.**

This definition requires that the partition element containing the observable universe is unexceptional in the distribution $F(\rho_i)$. In fact it requires only that $\rho_1$ not be so exceptional that the set of more extreme members has measure zero.

If the multiverse satisfies this Principle of Mediocrity, $S$ must be infinite. To see this assume $S$ is finite and choose $P_1$ large enough so that it contains all elements of $S$. Then $\rho_1 \neq 0$ and $\rho_i = 0$ for $i \neq 1$, hence $F(\rho_1) = 1$. This contradicts the definition, and the assumption that $S$ is finite is false.

The Principle of Mediocrity also implies that the $S$ has non-zero measure. Since $P_1$ is finite and contains at least one observer-supporting member, $\rho_1 \neq 0$. By definition $F(\rho_1) \neq 0$ or $1$, hence $P(\rho_j \leq \rho_1) = A$ with $A \neq 0$ or $1$. A lower bound $L$ on the measure of $S$ is given by $L = A.0 + (1-A).\rho_1 \neq 0$.

The simplest assumption concerning the composition of $M$ is a steady state infinite multiverse with no natural dimension or timescale. Individual elements of the ensemble may exist for different periods, but the overall composition does not vary. In a steady state multiverse a suitable observer reference class is the set of all observers at any epoch. For a non-steady multiverse it is necessary to include observers at all epochs, but this distinction is not critical for the Bayesian argument.

Returning to the calculation of the credence of multiverse and non-multiverse cosmologies, consider hypotheses $H_1$, $H_2$ and evidence $E$ as follows:

$H_1$: Physical laws and cosmological parameters are homogeneous and permit observers.

$H_2$: Physical laws and cosmological parameters occur in an *infinite* ensemble $M$ with subset $S$ of non-zero measure that permit observers.

$E$: The observed universe permits observers.

Again cosmology $H_1$ is fine tuned and cosmology $H_2$ is not, and the ratio of prior probabilities $P(H_1) / P(H_2) = F$ where $F$ is the fine tuning parameter. Applying SSA with the reference class of all observers in the ensemble, the conditional probability of $E$ under $H_1$ or $H_2$ is unity as all observers in $H_2$ are in elements of $S$. Freak observers may be neglected due to their extreme improbability, and the fact that the measure of $S$ is non-zero. Then by Bayes' theorem:

$$\frac{P(H_1/E)}{P(H_2/E)} = \frac{P(E/H_1)P(H_1)}{P(E/H_2)P(H_2)} = F$$

Thus fine tuning favors a heterogenous multiverse over a fine tuned homogeneous universe, and the strength of this inference is given by the fine tuning parameter.

## 6. Conclusions and Implications

The result here rests on three assumptions. Firstly, that probability theory may be applied to the credence of hypotheses, and the Self Sampling Assumption is a valid way to treat observer selection effects. Secondly, that a homogeneous or heterogeneous cosmos are *a priori* equally likely, but fine tuning biases the prior probability to a heterogeneous multiverse. Finally, the multiverse obeys a Principle of Mediocrity such that a section containing the observer is unexceptional if sampled on a sufficiently large scale.

The first assumption is widely accepted, the second is strongly supported by theory and observations, and the third is plausible but cannot be experimentally validated.

The approach followed is to define two classes of cosmologies; one in which the physical laws are homogeneous and permit observers, and another in which an ensemble of domains spans the possible laws. Two hypotheses are then defined according to whether the cosmos conforms to one class or the other. Bayes' theorem is applied to determine the relative credence of the two hypotheses.

Note that these hypotheses do not include all possible cosmologies. Models in which the entire multiverse is fine tuned are not addressed as they offer no resolution of the fine tuning problem. Also, cosmologies involving the agency of intelligence are excluded by the tacit assumption that elements of the multiverse ensemble are generated by natural processes, and thus without planning. Any conscious design mechanism makes the fine tuning problem moot, but at the cost of introducing the larger problem of explaining the existence and mechanism of primordial intelligence.

Note that the intercession of intelligence does not imply supernatural phenomena. It has been proposed that life may advance to the extent that it achieves the capability to create universes and select their properties to favor the subsequent emergence of life, indicating a process akin to universal natural selection (Harrison, 1995).

The present analysis does not address these speculations. It indicates only the relative likelihood of two hypotheses. The conclusion may be stated as follows: If the observable universe is fine tuned, a heterogenous multiverse is more likely than a homogeneous universe.

The strength of the inference favoring the multiverse is given by the degree of fine tuning. It may prove difficult to determine this parameter until a full theory of physics is available, which allows calculation of the distribution of cosmological properties. If the chance that these properties support observers is small, indicating strong fine tuning, this constitutes evidence favoring a heterogenous multiverse.

The multiverse hypothesis may never be subjected to experimental falsification. While the idea has substantial explanatory power, it may yield no direct observational consequences. This suggests it may be regarded as a discipline of philosophy rather than science. Nevertheless, there are sound reasons to suggest it may be true.

Perhaps the most significant implication of the multiverse concept is that it highlights a limitation of science itself. If the quest for a full theory of physics succeeds, and explains all phenomena in the observable universe, it may be only a special case from a vast ensemble of such theories. The existence and properties of such other realms may remain matters of speculation, forever beyond the reach of experiment or observation.